\documentclass[aps,preprint,nofootinbib,floatfix]{revtex4-2}
\usepackage[utf8]{inputenc}

\usepackage[title]{appendix}
\usepackage[colorlinks,
linkcolor=blue,
filecolor=blue,
anchorcolor=blue,
urlcolor=blue,
citecolor=blue,
bookmarks=false,
]{hyperref}
\usepackage{graphicx}
\usepackage{subfigure}
\usepackage{amssymb}
\usepackage{amsmath}
\usepackage{bm}
\usepackage{dsfont}
\usepackage{booktabs}
\usepackage{xcolor}
\allowdisplaybreaks[4]
\begin{document}
\title{\bf\boldmath Probing neutrino-light-quark effective scalar interactions from neutrino masses}
\author{Feng-Zhi Chen}
\author{Junlin Huang}
\author{Fanrong Xu}
\email{fanrongxu@jnu.edu.cn}

\affiliation{\it Department of Physics, College of Physics and Optoelectronic Engineering, Jinan University, Guangzhou 510632, P.R. China}


\begin{abstract}

In this work, we use neutrino masses as a probe of the neutrino-light-quark effective scalar interactions. It is found that neutrinos can acquire masses not only from the usual light quark loop corrections but also from the light quark condensates. The latter contribution has been overlooked in the literature. We show that both contributions are comparable for operators involving $u$ and $d$ quarks, while quark loop corrections dominate for operators involving the $s$ quark. Using the low-energy effective field theory extended with light right-handed neutrinos and matching to chiral perturbation theory, we systematically analyze these contributions, deriving constraints on the corresponding Wilson coefficients from neutrino mass bounds, coherent elastic neutrino-nucleus scattering, and light pseudoscalar meson invisible decays. Our analysis shows that electron neutrino mass measurements provide the most stringent constraints on these scalar couplings, significantly improving upon limits from other observables. The results highlight the importance of including both perturbative and nonperturbative contributions in complete phenomenological analyses of neutrino mass generation mechanisms.

\end{abstract}

\maketitle
\newpage

\section{Introduction}
\label{sec:intro}

In the Standard Model (SM), neutrinos are massless due to the absence of right-handed neutrinos or a Higgs triplet. However, the discovery of neutrino oscillations has demonstrated that neutrinos possess small but nonzero masses~\cite{Super-Kamiokande:1998kpq,SNO:2002tuh,KamLAND:2002uet}. Consequently, neutrino masses must originate from new physics (NP) beyond the SM. Investigating the origin of the smallness of neutrino masses provides a promising avenue for exploring the nature of NP.

Experimentally, the most recent terrestrial constraint on the neutrino mass from tritium $\beta$ decay comes from the KATRIN experiment, which sets an upper limit of $m_{\nu_e}\equiv\sqrt{\sum_i|U_{ei}|^2m_i^2}<0.45$~eV at $90\%$ confidence level (CL)~\cite{KATRIN:2024cdt}. Additionally, stringent bounds on the effective Majorana mass, $m_{ee}<28\text{--}122$~meV, are derived from the nonobservation of neutrinoless double beta decay by KamLAND-Zen~\cite{KamLAND-Zen:2024eml}. Cosmological observations of the cosmic microwave background also provide a model-dependent upper bound on the sum of neutrino masses, $\sum_i m_{\nu_i}<0.12$~eV ($95\%$ CL)~\cite{Planck:2018vyg}. In all cases, neutrino masses are at least 6 orders of magnitude smaller than those of all other SM fermions, suggesting that neutrinos acquire mass through a mechanism distinct from the Higgs mechanism~\cite{Higgs:1964pj,Englert:1964et}. 

This paper focuses on constraining the effective couplings of neutrino-light-quark scalar interactions using neutrino mass bounds. We assume that neutrinos acquire mass via scalar four-fermion effective interactions involving neutrinos and light quarks ($u,d,s$). Within this framework, we identify two distinct contributions to neutrino masses: (i) perturbative contributions from light quark loop corrections and (ii) nonperturbative contributions from light quark condensates. While neutrinos could also gain mass from scalar interactions involving heavy quarks or charged leptons via loop diagrams, we focus on light quarks due to their unique property: Below the chiral symmetry-breaking scale ($\mu_\chi\sim 1$~GeV), light quarks not only contribute perturbatively, but also develop a vacuum condensate, providing an additional contribution to the neutrino mass~\cite{Thomas:1992hf,Babic:2019zqu,Scholer:2024muv}. Moreover, the same scalar operators contributing to neutrino masses also contribute to low-energy processes such as coherent elastic neutrino-nucleus scattering (CE$\nu$NS) and invisible decays of light pseudoscalar mesons.\footnote{While these scalar operators may modify neutrino oscillations (see, e.g., Refs~\cite{Ge:2018uhz,Denton:2022pxt,Denton:2024upc}), such effects are expected to be subdominant and are thus excluded from our phenomenological analysis.} Thus, constraints derived from neutrino masses have significant implications for these processes.

It is noteworthy that previous studies have considered only the perturbative contribution, overlooking the nonperturbative effect~\cite{Prezeau:2004md,Ito:2004sh,Han:2020pff}. We demonstrate that both contributions are comparable for operators involving $u$ and $d$ quarks, whereas the perturbative contribution dominates for operators involving the $s$ quark. We discuss both Dirac and Majorana neutrino cases, as the nature of neutrinos remains unknown. The analysis is performed using the low-energy effective field theory extended with right-handed neutrinos $N_R$ (LNEFT)~\cite{Jenkins:2017jig,Jenkins:2017dyc,Chala:2020vqp,Li:2020lba}, and we include renormalization group (RG) running effects. The nonperturbative contributions are obtained by matching LNEFT operators onto chiral perturbation theory ($\chi$PT) operators.

The paper is organized as follows. Section~\ref{sec:EFT} introduces the scalar interactions in the effective field theories, including LNEFT, RG running effects, and the matching to $\chi$PT. Section~\ref{sec:obs} presents the expressions for relevant observables: neutrino masses, CE$\nu$NS, and light pseudoscalar meson invisible decays. Section~\ref{sec:num} provides a numerical analysis comparing the perturbative and nonperturbative contributions to neutrino masses and constrains the Wilson coefficients from various observables. Our conclusions are given in Sec.~\ref{sec:con}.

\section{Scalar interactions in effective field theories}
\label{sec:EFT}

In this section we will first describe all the relevant scalar effective interactions in the Lagrangian of LNEFT at scale $\mu_\chi\sim 1$~GeV, from which one can calculate the perturbative contribution to the neutrino masses. We subsequently match the LNEFT operators onto the $\chi$PT operators to obtain the nonperturbative contribution to the neutrino masses.

\subsection{Scalar interactions in LNEFT}\label{subsec:LNEFT}

LNEFT is an effective theory valid below the electroweak scale ($\mu_\mathrm{EW}$), invariant under Lorentz transformation and $SU(3)_C \times U(1)_{em}$ gauge symmetries. Its degrees of freedom include all SM fields except the heavy $W$, $Z$, $h$, and $t$, along with light right-handed neutrinos $N_R$~\cite{Jenkins:2017jig,Jenkins:2017dyc,Chala:2020vqp,Li:2020lba}. We consider only the scalar four-fermion operators involving neutrinos and light quarks.

Depending on lepton number conservation, these operators are categorized into two classes: lepton-number-conserving (LNC) and lepton-number-violating (LNV). For Dirac neutrinos, the relevant LNC Lagrangian is
\begin{align}\label{eq:LNCLNEFT}
\mathcal{L}_\mathrm{LNEFT}^\mathrm{LNC}\supset &L_{\substack{\nu Nu\\prst}}^{S,RR}(\bar{\nu}_{Lp}N_{Rr})(\bar{u}_{Ls}u_{Rt})+L_{\substack{\nu Nu\\prst}}^{S,RL}(\bar{\nu}_{Lp}N_{Rr})(\bar{u}_{Rs}u_{Lt})\notag\\
&+L_{\substack{\nu Nd\\prst}}^{S,RR}(\bar{\nu}_{Lp}N_{Rr})(\bar{d}_{Ls}d_{Rt})+L_{\substack{\nu Nd\\prst}}^{S,RL}(\bar{\nu}_{Lp}N_{Rr})(\bar{d}_{Rs}d_{Lt})+\mathrm{H.c.}\,,
\end{align}
where the coupling constant $L_i$ in front of each four-fermion operator is the Wilson coefficient, which is suppressed by a factor of $\mu^{-2}$, $L$ and $R$ denote separately the left and right chirality, and $p$, $r$, $s$, and $t$ denote the generation indices. Since both $\nu_L$ and $N_R$ carry a lepton number: $L(\nu_L)=L(N_R)=+1$, interactions governed by Eq.~\eqref{eq:LNCLNEFT} correspond to the LNC ($|\Delta L|=0$) processes. For Majorana neutrinos, the LNV Lagrangian is 
\begin{align}\label{eq:LNVLNEFT}
\mathcal{L}_\mathrm{LNEFT}^\mathrm{LNV}\supset &L_{\substack{\nu u\\prst}}^{S,LR}(\bar{\nu}^c_{Lp}\nu_{Lr})(\bar{u}_{Ls}u_{Rt})+L_{\substack{\nu u\\prst}}^{S,LL}(\bar{\nu}^c_{Lp}\nu_{Lr})(\bar{u}_{Rs}u_{Lt})\notag\\
&+L_{\substack{\nu d\\prst}}^{S,LR}(\bar{\nu}^c_{Lp}\nu_{Lr})(\bar{d}_{Ls}d_{Rt})+L_{\substack{\nu d\\prst}}^{S,LL}(\bar{\nu}^c_{Lp}\nu_{Lr})(\bar{d}_{Rs}d_{Lt})\notag\\
&+L_{\substack{N u\\prst}}^{S,RR}(\bar{N}^c_{Rp}N_{Rr})(\bar{u}_{Ls}u_{Rt})+L_{\substack{N u\\prst}}^{S,RL}(\bar{N}^c_{Rp}N_{Rr})(\bar{u}_{Rs}u_{Lt})\notag\\
&+L_{\substack{N d\\prst}}^{S,RR}(\bar{N}^c_{Rp}N_{Rr})(\bar{d}_{Ls}d_{Rt})+L_{\substack{N d\\prst}}^{S,RL}(\bar{N}^c_{Rp}N_{Rr})(\bar{d}_{Rs}d_{Lt})+\mathrm{H.c.}\,,
\end{align}
where the charge conjugation of a left-handed field $\psi_{L}^c=\mathcal{C}\bar{\psi}_{L}^T$ ($\mathcal{C}=i\gamma_2\gamma_0$) is actually a right-handed field, and vice versa. Since the antiparticle of a Majorana neutrino is the particle itself, operators in Eq.~\eqref{eq:LNVLNEFT} violate the lepton number by two units ($|\Delta L|=2$), interactions governed by which correspond to the LNV processes.

\subsection{RG running effects}
\label{subsec:RG}

Since there is a large mass gap between scales $\mu_\chi$ and $\mu$, one has to resume the large logarithms produced by the ratio of the two scales in the perturbative expansion. This can be obtained solving the relevant RG equations. Note that LNEFT is merely valid below $\mu_\mathrm{EW}$, at the range between $\mu_\mathrm{EW}$ and $\mu$; however, the SM gauge symmetries are still unbroken, and physics at this range should be described by the SM effective field theory ~\cite{Buchmuller:1985jz,Grzadkowski:2010es,Lehman:2014jma,Brivio:2017vri} extended with right-handed neutrinos (SMNEFT)~\cite{delAguila:2008ir,Aparici:2009fh,Bhattacharya:2015vja,Liao:2016hru,Liao:2016qyd,Bischer:2019ttk}. The Lagrangian of SMNEFT is invariant under Lorentz transformation and $SU(3)_C\times SU(2)_L\times U(1)_Y$ gauge symmetries, and the degrees of freedom of which are all the SM fields and the extended right-handed neutrinos $N_R$. To connect the physics at $\mu_\chi$ to the UV-complete model at $\mu$, a standard procedure is describing the low-energy physics in terms of LNEFT at $\mu_\chi$, evolving the LNEFT Wilson coefficients from $\mu_\chi$ up to $\mu_\mathrm{EW}$, matching LNEFT onto SMNEFT at $\mu_\mathrm{EW}$, running the SMNEFT Wilson coefficients from $\mu_\mathrm{EW}$ up to $\mu$, and matching SMNEFT to the UV-complete model at $\mu$. Nevertheless, as we are interested in only the neutrino mass generation mechanism and the neutrino low-energy physics, strictly executing such a procedure will be rather tricky and is out of the scope of our research. For simplicity, we neglect the RG evolutions for SMNEFT Wilson coefficients from $\mu$ down to $\mu_\mathrm{EW}$, and consider only the one-loop RG running effects from $\mu_\mathrm{EW}$ down to $\mu_\chi$ for LNEFT Wilson coefficients in this work. This treatment is reasonable, as the dominant running effect is mainly obtained from solving the QCD one-loop RG equations between $\mu_\mathrm{EW}$ and $\mu_\chi$, with a result by multiplying the Wilson coefficients at $\mu_\mathrm{EW}$ a factor of about 2. By the way, we also neglect the potential operator mixing that could appear in either LNEFT or SMNEFT, which is equal to assume that the contribution produced only by scalar operators.

Working in the $\overline{\mathrm{MS}}$ renormalization scheme, the one-loop QCD and QED RG equations for the scalar LNEFT Wilson coefficients read
\begin{align}\label{eq:LNEFTRGE}
\mu\frac{d}{d\mu}L_{\substack{i q'\\prqq}}^{S,AB}=-\left(6C_F\frac{\alpha_s}{4\pi}+6Q_q^2\frac{\alpha}{4\pi}\right)L_{\substack{i q'\\prqq}}^{S,AB}\,, \qquad  i=\nu, N, \nu N\,,
\end{align}
where $A,B=L,R$, $q'=u,d$ (when $q'=u$, $q=u$, and when $q'=d$, $q=d$ or $s$), $\alpha_s=g_s^2/4\pi$ and $\alpha=e^2/4\pi$ denote separately the coupling constants of strong and electromagnetic interactions, $Q_q$ is the electric charge for quark with $Q_u=+2/3$ and $Q_d=Q_s=-1/3$, and $C_F=(N_C^2-1)/2N_C=4/3$ with $N_C=3$ being the color number of QCD. The solution of Eq.~\eqref{eq:LNEFTRGE} is
\begin{align}\label{eq:RGsolution}
L_{\substack{i q'\\prqq}}^{S,AB}(\mu_1)=\left[\frac{\alpha_s(\mu_2)}{\alpha_s(\mu_1)}\right]^{-\frac{6C_F}{2\beta_0}}\left[\frac{\alpha(\mu_2)}{\alpha(\mu_1)}\right]^{-\frac{6Q_q^2}{2b_{0,e}}}\!L_{\substack{i q'\\prqq}}^{S,AB}(\mu_2)\,,
\end{align}
where $\beta_0=11-2n_f/3$ with $n_f$ being the number of active quarks between $\mu_1$ and $\mu_2$ and $b_{0,e}=-4(3n_\ell+4n_u+n_d)/9$ with $n_\ell$, $n_u$, and $n_d$ being, respectively, the numbers of active charged leptons, up-type quarks and down-type quarks between $\mu_1$ and $\mu_2$. Numerically, by quoting the input $\alpha_s(m_Z)=0.1180(9)$ and $\alpha=1/137.036$ from PDG~\cite{ParticleDataGroup:2024cfk}, one has
\begin{align}
L_{\substack{i q'\\prqq}}^{S,AB}(1~\text{GeV})\approx 2.07\,L_{\substack{i q'\\prqq}}^{S,AB}(m_Z)\,.
\end{align}
Clearly, the scalar Wilson coefficients get enhanced when running from a high scale down to a low scale. Note that the running effects of QED are negligible compared to that of QCD, which has also been pointed out in Ref.~\cite{Li:2020lba}.

\subsection{Matching LNEFT to $\chi$PT}
\label{subsec:chiPT}

For energy below $\mu_\chi\sim4\pi F_\pi$ with $F_\pi=92.3(1)$~MeV being the pion decay constant~\cite{Manohar:1983md}, the observed degrees of freedom are no longer the light quarks but the pseudoscalars ($\pi, K, \eta$), which can be well described by the so-called $\chi$PT. Therefore, to describe physics below $\mu_\chi$ with contributions from high energies, the LNEFT operators should be matched onto the ones of $\chi$PT. 

The $\chi$PT is a QCD low-energy effective field theory, in which the pseudoscalars are the Goldstone bosons corresponding to the spontaneous global chiral symmetries breaking: $SU(3)_L\times SU(3)_R\to SU(3)_V$. There are produced eight pseudoscalars, which can be written into the theory in a nonlinear way:
\begin{align}
U=\mathrm{exp}\left(i\frac{\Phi}{F_0}\right)\,,\qquad \Phi=\lambda^a\phi^a=\begin{pmatrix}
\pi^0+\frac{1}{\sqrt{3}}\eta & \sqrt{2}\pi^+ &\sqrt{2}K^+\\
\sqrt{2}\pi^- & -\pi^0+\frac{1}{\sqrt{3}}\eta & \sqrt{2}K^0\\
\sqrt{2}K^- &\sqrt{2}\bar{K}^0 & -\frac{2}{\sqrt{3}}\eta\\
\end{pmatrix}\,,
\end{align}
where $F_0$ is the pion decay constant in the chiral limit, $\phi^a$ are the Goldstone bosons, and $\lambda^a$ ($a=1,...,8$) are the Gell-Mann matrices satisfying the trace relation $\mathrm{Tr}(\lambda_a\lambda_b)=2\delta_{ab}$. Under $SU(3)_L\times SU(3)_R$ symmetries, the matrix $U$ transforms as $U\to RUL^\dagger$, with $L(R)$ being the representation of $ SU(3)_{L(R)}$. The chiral symmetries are broken by the quark vacuum expectation values or, equivalently, the quark condensates, $\langle 0|\bar{q}q|0\rangle$ ($q=u,d,s$), whose values can be determined through the lattice QCD calculations~\cite{McNeile:2012xh,Davies:2018hmw}. It is worth pointing out that the quark masses softly break the $SU(3)_L\times SU(3)_R$ symmetries which would result in nonzero masses pseudoscalars. 

Beside the pure QCD, one is usually dealing with interactions between quark bilinears and other fields, e.g., the neutrinos along with the accompanying Wilson coefficients in Eqs.~\eqref{eq:LNCLNEFT} and~\eqref{eq:LNVLNEFT}. These fields can be treated as the external sources (which are also referred as the spurion fields~\cite{Dekens:2018pbu}), which can be organized into the Lagrangian by employing the external field method~\cite{Gasser:1983yg,Gasser:1984gg,Cata:2007ns}:
\begin{align}\label{eq:ext}
\mathcal{L}=&\mathcal{L}_\mathrm{QCD}^0+\mathcal{L}_\mathrm{ext}\notag\\
           =&\mathcal{L}_\mathrm{QCD}^0+\bar{q}_L\gamma^\mu l_\mu q_L+\bar{q}_R\gamma^\mu r_\mu q_R+\bar{q}_L S q_R+\bar{q}_R S^\dagger q_L+\bar{q}_L \sigma^{\mu\nu}t_{\mu\nu} q_R+\bar{q}_R \sigma^{\mu\nu}t_{\mu\nu}^\dagger q_L\,,
 \end{align} 
where $q=(u,d,s)^T$ and $l_\mu$, $r_\mu$, $S$, and $t_{\mu\nu}$ are $3\times3$ Hermitian matrices in flavor space, which stand for the left-handed, right-handed, scalar, and tensor external sources, respectively. In writing down Eq.~\eqref{eq:ext}, we have endowed the external sources with the following chiral power counting:
\begin{align}\label{eq:counting}
l_\mu\sim \mathcal{O}(p)\,,\quad r_\mu\sim \mathcal{O}(p)\,,\quad S\sim \mathcal{O}(p^2)\,,\quad t_{\mu\nu}\sim \mathcal{O}(p^2)\,.
\end{align}
For convenience, the external sources can split into two components~\cite{Dekens:2018pbu}:
\begin{align}
l_\mu\mapsto l_\mu+\tilde{l}_\mu\,,\quad r_\mu\mapsto r_\mu+\tilde{r}_\mu\,,\quad S\mapsto S+\tilde{S}\,,\quad t_{\mu\nu}\mapsto t_{\mu\nu}+\tilde{t}_{\mu\nu}\,,
\end{align}
where $S$, $l_\mu$, $r_\mu$, and $t_{\mu\nu}$ in the right side of each arrow denote the quark mass matrix and the couplings to electromagnetic field $A_\mu$ in the SM:
\begin{align}\label{eq:split}
S\mapsto -M^\dagger\,,\quad l_\mu\mapsto -eQA_\mu\,,\quad r_\mu\mapsto -eQA_\mu\,,\quad t_{\mu\nu}\mapsto 0\,,
\end{align}
with $M=\mathrm{diag}(m_u,m_d,m_s)$ and $Q=\mathrm{diag}(2/3,-1/3,-1/3)$; whereas $\tilde{S}$, $\tilde{l}_\mu$, $\tilde{r}_\mu$, and $\tilde{t}_{\mu\nu}$ encode the contributions from NP. 

As we are interested in only the scalar interactions between quarks and neutrinos, we merely focus on the term $\bar{q}_L \tilde{S} q_R$ (and its Hermitian conjugate $\bar{q}_R \tilde{S}^\dagger q_L$) in Eq.~\eqref{eq:ext}. Concretely, the $\tilde{S}$ matrix in the flavor space can be written as
\begin{align}
\tilde{S}=\begin{pmatrix}
\tilde{S}_{uu} & \tilde{S}_{ud} & \tilde{S}_{us}\\
\tilde{S}_{du} & \tilde{S}_{dd} & \tilde{S}_{ds}\\
\tilde{S}_{su} & \tilde{S}_{sd} & \tilde{S}_{ss}
\end{pmatrix}\,.
\end{align}
Therefore, by matching $\bar{q}_L \tilde{S} q_R$ to the LNC LNEFT operators in Eq.~\eqref{eq:LNCLNEFT}, one has
\begin{align}
\tilde{S}_{uu}=&L_{\substack{\nu Nu\\pruu}}^{S,RR}(\bar{\nu}_{Lp}N_{Rr})+L_{\substack{\nu Nu\\pruu}}^{S,RL\ast}(\bar{N}_{Rr}\nu_{Lp})\,,\label{eq:LNCscalarex}\\[0.2cm]
\tilde{S}_{dd}=&L_{\substack{\nu Nd\\prdd}}^{S,RR}(\bar{\nu}_{Lp}N_{Rr})+L_{\substack{\nu Nd\\prdd}}^{S,RL\ast}(\bar{N}_{Rr}\nu_{Lp})\,,\\[0.2cm]
\tilde{S}_{ss}=&L_{\substack{\nu Nd\\prss}}^{S,RR}(\bar{\nu}_{Lp}N_{Rr})+L_{\substack{\nu Nd\\prss}}^{S,RL\ast}(\bar{N}_{Rr}\nu_{Lp})\,.
\end{align}
Similarly, by matching $\bar{q}_L \tilde{S} q_R$ to the LNV LNEFT operators in Eq.~\eqref{eq:LNVLNEFT}, one obtains
\begin{align}
\tilde{S}_{uu}=&L_{\substack{\nu u\\pruu}}^{S,LR}(\bar{\nu}_{Lp}^c\nu_{Lr})+L_{\substack{\nu u\\pruu}}^{S,LL\ast}(\bar{\nu}_{Lr}\nu_{Lp}^c)+L_{\substack{N u\\pruu}}^{S,RR}(\bar{N}_{Rp}^cN_{Rr})+L_{\substack{N u\\pruu}}^{S,RL\ast}(\bar{N}_{Rr}N_{Rp}^c)\,,\\[0.2cm]
\tilde{S}_{dd}=&L_{\substack{\nu d\\prdd}}^{S,LR}(\bar{\nu}_{Lp}^c\nu_{Lr})+L_{\substack{\nu d\\prdd}}^{S,LL\ast}(\bar{\nu}_{Lr}\nu_{Lp}^c)+L_{\substack{N d\\prdd}}^{S,RR}(\bar{N}_{Rp}^cN_{Rr})+L_{\substack{N d\\prdd}}^{S,RL\ast}(\bar{N}_{Rr}N_{Rp}^c)\,,\\[0.2cm]
\tilde{S}_{ss}=&L_{\substack{\nu d\\prss}}^{S,LR}(\bar{\nu}_{Lp}^c\nu_{Lr})+L_{\substack{\nu d\\prss}}^{S,LL\ast}(\bar{\nu}_{Lr}\nu_{Lp}^c)+L_{\substack{N d\\prss}}^{S,RR}(\bar{N}_{Rp}^cN_{Rr})+L_{\substack{N d\\prss}}^{S,RL\ast}(\bar{N}_{Rr}N_{Rp}^c)\,.\label{eq:LNVscalarex}
\end{align}

On the other hand, the $\chi$PT counterparts of $\bar{q}_L \tilde{S} q_R$ at the lowest chiral power order [$\mathcal{O}(p^{4})$] read~\cite{Dekens:2018pbu}
\begin{align}\label{eq:scalarMatch}
\bar{q}_{L}\tilde{S}q_{R}\to-2B_{0}\bigg[&\frac{1}{4}F_{0}^{2}\big\langle\tilde{S}U\big\rangle+L_{4}\big\langle D_{\mu}U^{\dagger}D^{\mu}U\big\rangle\big\langle\tilde{S}U\big\rangle+L_{5}\big\langle\tilde{S}UD_{\mu}U^{\dagger}D^{\mu}U\big\rangle \notag\\
 & +2L_{6}\langle U^{\dagger}\chi+\chi^{\dagger}U\rangle\langle\tilde{S}U\rangle-2L_{7}\langle U^{\dagger}\chi-\chi^{\dagger}U\rangle\langle\tilde{S}U\rangle+2L_{8}\langle\tilde{S}U\chi^{\dagger}U\rangle \notag\\
 & +H_{2}\langle\tilde{S}\chi\rangle\bigg]+\mathcal{O}(p^{6})\,,
\end{align}
where $\langle\cdots\rangle$ represents the trace in flavor space, $\chi=-2B_0S^\dagger$ with $B_0=-\langle 0|\bar{q}q|0\rangle/F_0^2$, and $L_i$ ($i=4,5,6,7,8$) and $H_2$ the low-energy coupling constants of the corresponding terms. It is clear from Eq.~\eqref{eq:scalarMatch} that this term is proportional to a constant $B_0$, which is actually connected to the quark condensates, and is, therefore, a nonperturbative effect. Expanding the first term on the right side of Eq.~\eqref{eq:scalarMatch} at the lowest energy, we will obtain an effective mass term for neutrinos; for more details, see Sec.~\ref{subsec:nm}. By the way, it is noteworthy that a similar procedure has also been applied to obtain the nonperturbative effects in other observables, e.g., the charged lepton and neutrino electric or magnetic dipole moments~\cite{Dekens:2018pbu,Chen:2022xkk}. 

\section{Observables}\label{sec:obs}

In this section, we will in detail represent the formulas for the relevant observables, which are expressed in terms of the LNEFT Wilson coefficients. These expressions will be used for the numerical analysis in the next section.

\subsection{Neutrino masses}\label{subsec:nm}

As introduced in Sec.~\ref{sec:intro}, the neutrinos acquire mass through two distinct ways: (i) the perturbative contribution from light quark loop corrections and (ii) the nonperturbative contribution from light quark condensates. In the following, we will separately display the detailed formulas for these two sources.

\begin{figure}[htb]
    \centering
        \includegraphics[width=0.328\linewidth]{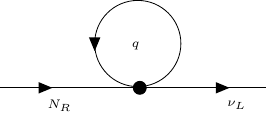}
        \includegraphics[width=0.328\linewidth]{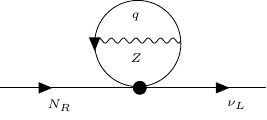}
        \includegraphics[width=0.328\linewidth]{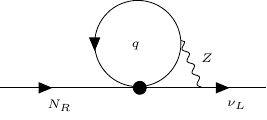}
    \vspace{-1.2cm}
    \caption{\small One- and two-loop Feynman diagrams contributing to the Dirac neutrino masses, where the blob denotes the neutrino-light-quark scalar effective interactions. Similar Feynman diagrams also apply to generate Majorana neutrino masses.}\label{fig:loop}
\end{figure}

The perturbative contribution is produced from the Feynman diagrams depicted in Fig.~\ref{fig:loop}.\footnote{Note that the scalar effective interactions should be produced at a scale above $\mu_\mathrm{EW}$, so that the heavy degree of freedom $Z$  can appear in the two-loop diagrams.} These types of Feynman diagrams were first calculated in Refs.~\cite{Prezeau:2004md,Ito:2004sh}, with an effective neutrino mass at one- and two-loop levels given, respectively, by
\begin{align}
\delta m_{\nu_\mathrm{eff}}^{(1)}\approx &\frac{1}{2}N_C L_{\substack{i q'\\prqq}}^{S,AB}\frac{m_q^3}{(4\pi)^2}\log\frac{\mu^2}{m_q^2}\,,\label{eq:1loop}\\[0.2cm]
\delta m_{\nu_\mathrm{eff}}^{(2)}\approx &\frac{1}{2}g^2N_C L_{\substack{i q'\\prqq}}^{S,AB}\frac{m_qm_Z^2}{(4\pi)^4}\left(\log\frac{\mu^2}{m_Z^2}\right)^2\,, \label{eq:2loop}
\end{align}
where $g$ is the $SU(2)_L$ gauge coupling constant. It is found that the ratio of the one-loop result to the two-loop result is $\sim 50m_q^2/m_Z^2$, so the one-loop contribution is negligible compared to the two-loop one; see more details about this aspect in Refs.~\cite{Prezeau:2004md,Ito:2004sh}. Note that, since $N_R$ is an $SU(2)_L\times U(1)_Y$ singlet, only the first two diagrams have contribution to the pure right-handed neutrino mass, yet for the sake of uniformity we keep the same formalism for all the neutrino mass terms. The relations between the Wilson coefficients of this work and the ones in Ref.~\cite{Prezeau:2004md} are
\begin{align}
L_{\substack{\nu N q'\\prqq}}^{S,RR}=&2G_F(a_{\mathrm{S};pr}^{qq}+a_{\mathrm{SP};pr}^{qq})\,, &  L_{\substack{\nu N q'\\prqq}}^{S,RL}=&2G_F(a_{\mathrm{S};pr}^{qq}-a_{\mathrm{SP};pr}^{qq})\,,\\
L_{\substack{\nu q'\\prqq}}^{S,LR}=&2G_F(a_{\mathrm{S};pr}^{qq}-a_{\mathrm{SP};pr}^{qq})\,, &  L_{\substack{\nu q'\\prqq}}^{S,LL}=&2G_F(a_{\mathrm{S};pr}^{qq}+a_{\mathrm{SP};pr}^{qq})\,,
\end{align}
where $G_F$ is the Fermi constant and the subscripts S and SP refer to scalar and pseudoscalar, respectively. In the numerical analysis, we consider only the predominant two-loop result, and the total perturbative contribution is obtained by multiplying Eq.~\eqref{eq:2loop} with the RG running effect in Eq.~\eqref{eq:RGsolution}.

The nonperturbative contribution, on the other hand, can be obtained from the first term in Eq.~\eqref{eq:scalarMatch}, which corresponds to the contribution from the quark condensate. To see this, we substitute the matching relations in Eqs.~\eqref{eq:LNCscalarex}--\eqref{eq:LNVscalarex} into their $\chi$PT counterparts~\eqref{eq:scalarMatch} and take only the vacuum state of the theory, i.e., the $3\times3$ identity matrix of the $U$ matrix ($U=\mathds{1}+i\Phi/F_0+\cdots$), to obtain the following effective mass term:
\begin{align}\label{eq:nonper}
\delta m_{\nu_\mathrm{eff}}^{qc}=-\frac{1}{2}B_0F^2_0\sum_{q}\tilde{S}_{qq}=\frac{1}{2}\langle 0|\bar{q}q|0\rangle \sum_{q}L_{\substack{i q'\\prqq}}^{S,AB}\,.
\end{align}
The numerical value of $\langle 0|\bar{q}q|0\rangle$ will be quoted from the results of lattice QCD~\cite{McNeile:2012xh,Davies:2018hmw}. The total neutrino effective mass is the sum of both the perturbative and nonperturbative contributions. We will numerically compare these two effects in Sec.~\ref{sec:num}.

\subsection{CE$\nu$NS}

In a CE$\nu$NS process, the low-energy (anti)neutrinos can coherently couple to protons and neutrons of a given nucleus $\mathcal{N}$, the cross section of which could significantly get enhanced and, therefore, allows for much smaller detectors (see, e.g., the recent proceedings on the CE$\nu$NS measurement in the  CONUS$+$ experiment~\cite{Ackermann:2025obx}). For this, it has been theoretically proposed long ago to use the CE$\nu$NS to probe the weak neutral current~\cite{Freedman:1973yd}. Experimentally, the CE$\nu$NS process was observed by the COHERENT experiment first in cesium iodide~\cite{COHERENT:2017ipa} and later confirmed in argon~\cite{COHERENT:2020iec} and germanium~\cite{COHERENT:2025vuz}, all are consistent with the SM predictions. One, therefore, can estimate impact on the CE$\nu$NS with influence from various NP sources~\cite{Lindner:2016wff,Chang:2020jwl,Li:2020lba}. As the same LNEFT scalar operators contributing to neutrino masses can also contribute to the CE$\nu$NS, we will use the results of the Wilson coefficients constrained from the former to predict their impacts on the latter.  

The incoming (anti)neutrinos in a CE$\nu$NS process can be definitely selected, but the outgoing missing energies cannot be measured and might be either the left- or right-handed neutrinos. This allows us to set up a limit on the NP parameters that could have a contribution to the process. In the presence of scalar interactions, the differential cross section for the scattering $\overset{(-)}{\nu}\mathcal{N}\to X\mathcal{N}$ with $X\in\{\overset{(-)}{\nu},\overset{(-)}{N}\}$ can be written as~\cite{Lindner:2016wff,Coloma:2017egw,Chang:2020jwl,Li:2020lba}
\begin{align}
\frac{d\sigma}{dT}=\frac{G_F^2M}{4\pi}\left[\xi_S^2\frac{E_r}{E_r^\mathrm{max}}+\xi_V^2\left(1-\frac{E_r}{E_r^\mathrm{max}}\right)\right]\,,
\end{align}
where $M$ is the mass of the given nucleus, $E_r$ denotes recoil nucleus kinetic energy with maximal value $E_r^\mathrm{max}=2E_\nu^2/(M+E_\nu^2)\simeq 2E_\nu^2/M$ with $E_\nu$ being the energy of the incoming (anti)neutrinos, and $\xi_S$ and $\xi_V$  define the effective parameters describing the neutrino-nucleus interactions for scalar and vector currents, respectively. The SM contributes only to
the vector parameter $\xi_V$, with
\begin{align}
\xi_{V,\mathrm{SM}}^2=[\mathbb{N}-(1-4\sin^2\theta_W)\mathbb{Z}]^2F^2(q^2)
\end{align}
where $\mathbb{Z}$ and $\mathbb{N}$ label separately the proton and neutron numbers of nucleus $\mathcal{N}$, and $F(q^2)=F_p(q^2)=F_n(q^2)$ denotes the Helm form factor of the nucleus with the coherent limit $F(q^2\to 0)\simeq 1$~\cite{Helm:1956zz}. The parameter $\xi_S$ encodes the contributions from the scalar effective interactions, according to different participating neutrinos, one has~\cite{Lindner:2016wff,Chang:2020jwl,Li:2020lba}
\begin{align}\label{eq:xiS}
\xi_S^2=\begin{cases}
\sum\limits_{j;pr}\Big|\frac{1}{\sqrt{2}G_F}\sum\limits_{q}(L_{\substack{iq'\\prqq}}^{S,RR}+L_{\substack{iq'\\prqq}}^{S,RL})\left(\mathbb{Z}_j\frac{m_p}{m_q}f_{T_q}^p+\mathbb{N}_j\frac{m_n}{m_q}f_{T_q}^n\right)\Big|^2F^2(q^2)\,, & \text{for}~i=\nu N\\[0.2cm]
\sum\limits_{j;pr}\Big|\frac{\sqrt{2}}{G_F}\sum\limits_{q}(L_{\substack{iq'\\prqq}}^{S,LR}+L_{\substack{iq'\\prqq}}^{S,LL})\left(\mathbb{Z}_j\frac{m_p}{m_q}f_{T_q}^p+\mathbb{N}_j\frac{m_n}{m_q}f_{T_q}^n\right)\Big|^2F^2(q^2)\,, & \text{for}~i=\nu
\end{cases}\,,
\end{align}
where the subscript $j$ sums over the nucleus of a given molecule that participates the scattering, $m_p$ and $m_n$ denote, respectively, the mass of proton and neutron, and $f_{T_q}^{p,n}$ are the nucleon form factors for the scalar currents, with numeric values~\cite{Belanger:2008sj,Belanger:2018ccd}
\begin{align}
\begin{aligned}
f_{T_u}^{p}=&0.0153\,,\qquad & f_{T_d}^{p}=&0.0191\,,\qquad &  f_{T_s}^{p}=&0.0447\,,\\
f_{T_u}^{n}=&0.0110\,,\qquad & f_{T_d}^{n}=&0.0273\,,\qquad &  f_{T_s}^{n}=&0.0447\,.
\end{aligned}
\end{align}

\subsection{Light pseudoscalar meson invisible decays}

In the SM, the electric neutral light pseudoscalar ($\pi^0$, $\eta$, and $\eta'$) which consists of light quarks can decay into a neutrino pair via exchanging a $Z$ gauge boson. The SM predictions for these decay rates are highly suppressed, and only the upper limits were obtained from experiments, which at $90\%$ CL,
\begin{align}
\begin{aligned}
\mathcal{B}(\pi^0\to\mathrm{invisible})<&4.4\times10^{-9}~\text{\cite{NA62:2020pwi}}\,,\\
\mathcal{B}(\eta\to\mathrm{invisible})<&1.1\times10^{-4}~\text{\cite{NA64:2024mah}}\,,
\\
\mathcal{B}(\eta'\to\mathrm{invisible})<&2.1\times10^{-4}~\text{\cite{NA64:2024mah}}\,,
\end{aligned}
\end{align}
where the missing energy are supposed to be carried by the invisible neutrinos. As the missing energy may be mimicked by other exotic light degrees of freedom, the aforementioned light pseudoscalar meson invisible decays can be also used to set bounds on the potential NP parameters. 

In the presence of scalar interactions, the decay rate of a pseudoscalar meson $P$ (where $P\in\{\pi^0,\eta,\eta'\}$) invisible decay reads~\cite{Li:2020lba}
\begin{align}\label{eq:inv}
\mathcal{B}(P\to\mathrm{invisible})=\frac{\tau_Pm_P}{16\pi}\sum_{q;pr}\Bigg[&2\left|\frac{h_P^q}{4m_q}\left(L_{\substack{\nu Nq'\\prqq}}^{S,RR}-L_{\substack{\nu Nq'\\prqq}}^{S,RL}\right)\right|^2\left(1-\frac{m_{N_R}^2}{m_P^2}\right)^2\notag\\
&+\left|\frac{h_P^q}{2m_q}\left(L_{\substack{Nq'\\prqq}}^{S,RL}-L_{\substack{Nq'\\prqq}}^{S,RR}\right)\right|^2\left(1-2\frac{m_{N_R}^2}{m_P^2}\right)\left(1-4\frac{m_{N_R}^2}{m_P^2}\right)^{\frac{1}{2}}\notag\\
&+\left|\frac{h_P^q}{2m_q}\left(L_{\substack{\nu q'\\prqq}}^{S,LL}-L_{\substack{\nu q'\\prqq}}^{S,LR}\right)\right|^2\Bigg]\,,
\end{align}
where $m_{N_R}$, $m_q$, and $m_P$ denote the mass of the right-handed neutrino $N_R$, constituent quark $q$, and pseudoscalar meson $P$, respectively, and $h_P^q$ the decay constant which parametrized the hadronization of $P$ via a pseudoscalar current transition into vacuum:
\begin{align}
\langle 0|\bar{q}\gamma_5 q|P\rangle=-i\frac{h_P^q}{2m_q}\,.
\end{align}
Especially,
\begin{align}
\begin{aligned}
h_\pi^u=&m_\pi^2f_\pi^u\,,& h_\pi^d=&m_\pi^2f_\pi^d\,,& h_\pi^s=&0\,,\\
h_\eta^u=&\frac{c_\phi}{\sqrt{2}} h_q\,,& h_\eta^d=&\frac{c_\phi}{\sqrt{2}} h_q\,,& h_\eta^s=&-s_\phi h_s\,,\\
h_{\eta'}^u=&\frac{s_\phi}{\sqrt{2}} h_q\,,& h_{\eta'}^d=&\frac{s_\phi}{\sqrt{2}} h_q\,,& h_{\eta'}^s=&c_\phi h_s\,,
\end{aligned}
\end{align}
where $s_\phi\equiv\sin\phi$ and $c_\phi\equiv\cos\phi$ with $\phi=39.3^\circ$ being the $\eta-\eta'$ mixing angle in the Feldmann-Kroll-Stech scheme~\cite{Feldmann:1998vh,Feldmann:1998sh,Feldmann:1999uf} and
\begin{align}
\begin{aligned}
& h_q=f_q\left(m_\eta^2 c_\phi^2+m_{\eta^{\prime}}^2 s_\phi^2\right)-\sqrt{2} f_s\left(m_{\eta^{\prime}}^2-m_\eta^2\right) s_\phi c_\phi, \\
& h_s=f_s\left(m_{\eta^{\prime}}^2 c_\phi^2+m_\eta^2 s_\phi^2\right)-\frac{f_q}{\sqrt{2}}\left(m_{\eta^{\prime}}^2-m_\eta^2\right) s_\phi c_\phi
\end{aligned}
\end{align}
with $f_\pi^u=-f_\pi^d=F_\pi$, $f_q=1.51F_\pi$, and $f_s=1.90F_\pi$. Note that in the SM, only vector operators which arise from exchanging a $Z$ gauge boson contribute to the light pseudoscalar meson decays. However, due to that their contributions to the decay are helicity suppressed, such contributions are negligible~\cite{Li:2020lba}.

\section{Numerical analysis}
\label{sec:num}

In this section, we will first numerically compare the perturbative and nonperturbative contributions to the neutrino mass, then constrain and compare the LNEFT Wilson coefficients from the neutrino masses, CE$\nu$NS, and ($\pi^0$, $\eta$, $\eta'$) pseudoscalar meson invisible decays, and finally use the results obtained from neutrino mass bounds as inputs to predict the light pseudoscalar meson decay rates. For convenience, the numerical values for all inputs (and the corresponding references) used throughout this work are collected in Table~\ref{tab:input}.

\begin{table}[tbh!]
\centering
\tabcolsep 0.15in
\begin{tabular}{|l|l|}
\hline
  $G_F = 1.1663788 \times 10^{-5}~\mathrm{GeV}^{-2}$  \hfill\cite{ParticleDataGroup:2024cfk}  &
  $\alpha_s(m_Z) = 0.1180(9)$  \hfill\cite{ParticleDataGroup:2024cfk} \\
$\alpha=1/137.035999084(21)$   \hfill\cite{ParticleDataGroup:2024cfk} & 
   $\sin^2\theta_W=0.231 29(4)$  
\hfill  \cite{ParticleDataGroup:2024cfk} \\
  $m_{\pi^0}=134.9768(5)~\mathrm{MeV}$
  \hfill\cite{ParticleDataGroup:2024cfk}  &
 $\tau_{\pi^0}=8.43(13)\times10^{-17}~\mathrm{s}$
  \hfill\cite{ParticleDataGroup:2024cfk}  \\
  $m_{\eta} = 547.862(17)~\mathrm{MeV}$            
\hfill \cite{ParticleDataGroup:2024cfk} &
 $\Gamma_{\eta} = 1.31(5)~\mathrm{keV}$                
\hfill\cite{ParticleDataGroup:2024cfk} \\
$m_{\eta'} = 957.78(6)~\mathrm{MeV}$                
\hfill\cite{ParticleDataGroup:2024cfk} &
$\Gamma_{\eta'} = 0.188(6)~\mathrm{MeV}$        
 \hfill\cite{ParticleDataGroup:2024cfk} \\
  $m_{u} = 2.16(4)~\mathrm{MeV}$  \hfill\cite{ParticleDataGroup:2024cfk}    &
  $m_{d} = 4.70(4)~\mathrm{MeV}$
  \hfill\cite{ParticleDataGroup:2024cfk} \\
  $m_{s} = 93.5(5)~\mathrm{MeV}$  \hfill\cite{ParticleDataGroup:2024cfk}   &
$m_{c} = 1.2730(28)~\mathrm{GeV}$  \hfill\cite{ParticleDataGroup:2024cfk}   \\
  $m_{b} = 4.183(4)~\mathrm{GeV}$ \hfill\cite{ParticleDataGroup:2024cfk} &
 $m_Z = 91.1880(20)~\mathrm{GeV}$  
\hfill  \cite{ParticleDataGroup:2024cfk} \\
  $\langle0|\bar{u}u|0\rangle^{\overline{\mathrm{MS}}}(2~\mathrm{GeV}) = -(283(2)~\mathrm{MeV})^3$           
   \hfill\cite{McNeile:2012xh}   &
 $\langle0|\bar{d}d|0\rangle^{\overline{\mathrm{MS}}}(2~\mathrm{GeV}) = -(283(2)~\mathrm{MeV})^3$           
   \hfill\cite{McNeile:2012xh}    \\
 $\langle0|\bar{s}s|0\rangle^{\overline{\mathrm{MS}}}(2~\mathrm{GeV}) = -(296(11)~\mathrm{MeV})^3$           
   \hfill\cite{Davies:2018hmw} &             
 $F_{\pi} = 92.3(1)~\mathrm{MeV}$   \hfill\cite{Manohar:1983md} \\ 
\hline
\end{tabular}
\caption{\small The numerical values for the relevant inputs
 used in our numerical analysis.}
\label{tab:input}
\end{table}

\subsection{Perturbative contribution versus nonperturbative contribution}

As demonstrated in Sec.~\ref{subsec:nm}, the LNEFT neutrino-light-quark scalar operators have two contributions to the effective neutrino masses: one is the perturbative contribution arising from loop corrections, and the other is the nonperturbative contribution originating from the quark condensates. Using inputs from Table~\ref{tab:input}, the total effective neutrino mass for a single nonzero Wilson coefficient is
\begin{align}\label{eq:mnueff}
\delta m_\nu^\mathrm{eff}=\begin{cases}
(0.022_\text{loop}-0.011_{qc})L_{\substack{i q'\\prqq}}^{S,AB}\cdot\mathrm{GeV}^{3}\,,& \text{for}~q=u\\
(0.047_\text{loop}-0.011_{qc})L_{\substack{i q'\\prqq}}^{S,AB}\cdot\mathrm{GeV}^{3}\,,& \text{for}~q=d\\
(0.944_\text{loop}-0.013_{qc})L_{\substack{i q'\\prqq}}^{S,AB}\cdot\mathrm{GeV}^{3}\,, & \text{for}~q=s
\end{cases}\,,
\end{align}
where ``loop'' and ``$qc$'' above indicate the contribution induced from loop corrections and quark condensates, respectively. Obviously, we can conclude from Eq.~\eqref{eq:mnueff} that both contributions are comparable for $u$ and $d$ quarks, while the perturbative contribution dominates for $s$ quarks. Therefore, in the former case, both contributions should be taken into account, while we need to consider only the perturbative contribution in the latter case for which the nonperturbative contribution is negligible. Note that only the perturbative contribution has been considered in Ref.~\cite{Prezeau:2004md}, so their results should be refined. Besides, the updated bounds on the neutrino masses also demand renewed constraints on the scalar couplings.

It is important to note that the bounds derived in Eq.~\eqref{eq:mnueff} rely on a naturalness argument. We require that the contribution to the neutrino mass induced by the neutrino-light-quark scalar effective interactions does not exceed the experimentally observed neutrino mass. Mathematically, it is possible for these effective couplings to be significantly larger if the effective operator contributions are cancelled by a dimension-5 Weinberg operator (or other contributions) with opposite sign. Such a scenario would require a precise fine-tuning between unrelated model parameters. In the absence of a symmetry enforcing such a cancellation, we consider the naturalness bounds to be the most physically robust constraints.

While we have analyzed the neutrino mass generation within the framework of LNEFT, the embedding of these operators into a UV-complete model requires careful consideration of matching effects. In the context of SMNEFT, integrating out heavy scalars or vectors to generate dimension-6 operators typically induces a contribution to the dimension-5 Weinberg operator (or the Dirac Yukawa couplings and Majorana masses in models with right-handed neutrinos) at the one-loop matching level.

From a power-counting perspective, the matching coefficient for an operator of dimension $d$ scales as $\frac{C^{(d)}}{\mu^{d-4}}\sim \frac{y^n}{(16\pi^2)^k\mu^{d-4}}$, where $y$ represents the underlying UV couplings, $n$ denotes the coupling power depending on the specific Feynman diagram, and $k$ stands for the loop order. For a high UV scale $\mu$, the matching contribution to the neutrino mass from the dimension-5 operator generally dominates over the contribution from the dimension-6 operators, unless the underlying flavor symmetries of the UV theory suppress the dimension-5 operator while allowing the dimension-6 interactions.

\subsection{Constraints on the Wilson coefficients}

Since the same LNEFT Wilson coefficients simultaneously contribute to neutrino masses, CE$\nu$NS, and pseudoscalar meson invisible decays, different upper limits on these Wilson coefficients can be obtained from various observables. In this section, we will in detail compare the constraint capacity of different observables. 

Although our effective neutrino mass formulas apply to neutrinos of all flavors, we merely focus on the electron neutrino $\nu_e$, for which the constraints from the muon- ($\nu_\mu$) and tau- ($\nu_\tau$) neutrinos masses are still very poor, and the most stringent upper bound to date is from the $\nu_e$ mass. As the contributions from the scalar interactions of neutrinos and light quarks are only part of the corrections to the total neutrino masses, we have the following constraint:
\begin{align}
\delta m_{\nu}^\mathrm{eff}=\delta m_{\nu_\mathrm{eff}}^{(2)}+\delta m_{\nu_\mathrm{eff}}^{qc}<m_{\nu}^\mathrm{eff}\,.
\end{align}
For the effective Dirac $\nu_e$ mass, we will use the $90\%$ CL upper bound from KATRIN Collaboration~\cite{KATRIN:2024cdt}:
\begin{align}
m_{\nu_e}^\mathrm{eff}<0.45~\mathrm{eV}\,.
\end{align}
While for the effective Majorana electron neutrino mass, the strongest upper limit is taken from the bound on the $0\nu\beta\beta$ decay lifetime for xenon from the KamLAND-Zen experiment~\cite{KamLAND-Zen:2024eml}, which at $90\%$ CL is given by
\begin{align}
m_{ee}^\mathrm{eff}<28\text{--}122~\mathrm{meV}\,.
\end{align}
There has been no direct measurement on the light right-handed neutrino masses, so no available neutrino mass data can be utilized to constrain the relevant Wilson coefficients. However, the light pseudoscalar meson invisible decays can provide an indirect constraint on these Wilson coefficients, which, in turn, can be translated into the upper limit of correction to the mass of the right-handed neutrino.

In the following, we will compare the upper limits on the LNEFT Wilson coefficients obtained, respectively, from neutrino masses, CE$\nu$NS, and light pseudoscalar meson invisible decays. All the numerical values correspond to the LNEFT Wilson coefficients at the scale $\mu=1$~GeV and are obtained with the assumption that one coefficient is nonzero at a time. The constraints from the latter two have been done in Ref.~\cite{Li:2020lba} in the limit of massless right-handed neutrinos; we will quote their bounds from CE$\nu$NS based on the COHERENT measurement with cesium iodide~\cite{COHERENT:2017ipa} but use the updated constraints from the light pseudoscalar meson invisible decays. The comparisons of upper limits on LNC and LNV scalar LNEFT Wilson coefficients obtained from various observables are displayed in detail in Tables~\ref{tab:LNCupper} and~\ref{tab:LNVupper}, respectively. From these two tables, it is clear that neutrino mass bounds provide the most stringent constraints on both the LNC and LNV scalar LNEFT Wilson coefficients, surpassing those from CE$\nu$NS and light pseudoscalar meson invisible decays.

\begin{table}[htb!]
\centering
\tabcolsep 0.10in
\renewcommand\arraystretch{1.2}
\begin{tabular}{|c|c|c|c|c|c|}
\hline\hline
$L_i$~[GeV$^{-2}$] & $m_{\nu_e}^\mathrm{eff}$ & CE$\nu$NS & $\pi^0\to$inv & $\eta\to$inv & $\eta'\to$inv\\
\hline
$L_{\substack{\nu Nu\\eeuu}}^{S,RR(RL)}$ & $4.3\times10^{-8}$ & $7.6\times10^{-7}$ & $4.8\times10^{-7}$ & $8.5\times10^{-4}$ & $1.5\times10^{-2}$\\
\hline
$L_{\substack{\nu Nd\\eedd}}^{S,RR(RL)}$ & $1.2\times10^{-8}$ & $8.8\times10^{-7}$ & $1.1\times10^{-6}$ & $1.8\times10^{-3}$ & $3.2\times10^{-2}$ \\
\hline
$L_{\substack{\nu Nd\\eess}}^{S,RR(RL)}$ & $4.8\times10^{-10}$& $9.4\times10^{-6}$ & $\cdots$ & $5.5\times10^{-4}$ & $5.6\times10^{-3}$ \\
\hline\hline
\end{tabular}
\caption{\small Constraints on the upper limits of LNC scalar LNEFT Wilson coefficients from $m_{\nu_e}^\mathrm{eff}$, CE$\nu$NS, and light pseudoscalar meson invisible decays, respectively.}\label{tab:LNCupper}
\end{table}
\begin{table}[htb!]
\centering
\tabcolsep 0.10in
\renewcommand\arraystretch{1.2}
\begin{tabular}{|c|c|c|c|c|c|}
\hline\hline
$L_i$~[GeV$^{-2}$] & $m_{ee}^\mathrm{eff}$ & CE$\nu$NS & $\pi^0\to$inv & $\eta\to$inv & $\eta'\to$inv\\
\hline
$L_{\substack{\nu u\\eeuu}}^{S,LL(LR)}$ & $(1.3\text{--}5.8)\times10^{-9}$ & $3.8\times10^{-7}$ & $3.0\times10^{-7}$ & $6.0\times10^{-4}$ & $1.1\times10^{-2}$\\
\hline
$L_{\substack{\nu d\\eedd}}^{S,LL(LR)}$ & $(3.9\text{--}16.9)\times10^{-10}$ & $4.4\times10^{-7}$ & $6.7\times10^{-7}$ & $1.3\times10^{-3}$ & $2.3\times10^{-2}$ \\
\hline
$L_{\substack{\nu d\\eess}}^{S,LL(LR)}$ & $(1.5\text{--}6.6)\times10^{-11}$& $4.7\times10^{-6}$ & $\cdots$ & $3.9\times10^{-4}$ & $4.0\times10^{-3}$ \\
\hline
$L_{\substack{N u\\eeuu}}^{S,RR(RL)}$ & $\cdots$ & $\cdots$ & $3.0\times10^{-7}$ & $6.0\times10^{-4}$ & $9.2\times10^{-3}$ \\
\hline
$L_{\substack{N d\\eedd}}^{S,RR(RL)}$ & $\cdots$ & $\cdots$ & $6.7\times10^{-7}$ & $1.3\times10^{-3}$ & $2.0\times10^{-2}$ \\
\hline
$L_{\substack{N d\\eess}}^{S,RR(RL)}$ & $\cdots$ & $\cdots$ & $\cdots$ & $3.9\times10^{-4}$ & $4.0\times10^{-3}$ \\
\hline\hline
\end{tabular}
\caption{\small Constraints on the upper limits of LNV scalar LNEFT Wilson coefficients from $m_{ee}^\mathrm{eff}$, CE$\nu$NS, and light pseudoscalar meson invisible decays, respectively.}\label{tab:LNVupper}
\end{table}

From Table~\ref{tab:LNVupper}, we see that, among all of the light pseudoscalar meson invisible decays, the $\pi^0$ invisible decay provides the most stringent upper bound on the Wilson coefficients $L_{\substack{N u\\eeuu}}^{S,RR(RL)}$ and $L_{\substack{N d\\eedd}}^{S,RR(RL)}$, while the $\eta$ invisible decay imposes the most strict upper limit on $L_{\substack{N d\\eess}}^{S,RR(RL)}$. As discussed above, these upper limits can be translated into a correction to the mass of the light right-handed neutrino, with
\begin{align}
\delta m_{N_R}^\mathrm{eff}<\begin{cases}
6.6~\mathrm{eV}\,, & \text{from}~L_{\substack{N u\\eeuu}}^{S,RR(RL)}\\
48.2~\mathrm{eV}\,, & \text{from}~L_{\substack{N d\\eedd}}^{S,RR(RL)}\\
0.7~\mathrm{MeV}\,, & \text{from}~L_{\substack{N d\\eess}}^{S,RR(RL)}\\
\end{cases}\,.
\end{align}
Future more precise measurement on these light pseudoscalar meson invisible decays will improve the current upper bounds on the mass of the light right-handed neutrino.

\section{Conclusion}
\label{sec:con}

This work has systematically investigated the constraints on neutrino-light-quark scalar couplings from neutrino mass measurements, considering both perturbative loop corrections and nonperturbative quark condensate contributions. We have shown that, for operators involving $u$ and $d$ quarks, both contributions are comparable and must be included in complete analyses, while for $s$ quark operators, perturbative effects dominate.

By comparing constraints from various observables, we find that electron neutrino mass measurements provide the most stringent limits on the relevant LNEFT Wilson coefficients, significantly outperforming constraints from CE$\nu$NS and light pseudoscalar meson invisible decays. For Dirac neutrinos, the effective electron neutrino mass bound of 0.45~eV constrains the couplings to approximately $10^{-8}\text{--}10^{-10}$ GeV$^{-2}$, while for Majorana neutrinos, the more stringent $28\text{--}122$~meV bound from neutrinoless double beta decay provides constraints at the $10^{-9}\text{--}10^{-11}$ GeV$^{-2}$ level. It should be stressed that in deriving these constraints we have assumed the absence of accidental fine-tuned cancellations between the neutrino-light-quark scalar effective operator contributions and other contributions to the neutrino mass matrix, such as the dimension-5 Weinberg operator.

Furthermore, we have translated the constraints from light pseudoscalar meson invisible decays into limits on possible right-handed neutrino masses, obtaining bounds ranging from eV to MeV scales depending on the specific operator. Our results update previous analyses that considered only perturbative contributions and demonstrate the importance of a comprehensive treatment including both perturbative and nonperturbative effects.

The methodology developed in this work provides a framework for connecting UV-complete models with low-energy observables through effective field theory techniques, with potential applications to other beyond-Standard Model scenarios involving neutrino-mass generation mechanisms.

\section*{Acknowledgements}

This work is supported by NSFC under Grants No.~12475095 and~U1932104 and the 2024 Guangzhou Basic and Applied Basic Research Scheme Project for Maiden Voyage (2024A04J4190).

\bibliographystyle{apsrev4-1}
\bibliography{reference}

\end{document}